\documentclass[
   ,final            
  ]
  {aipproc}

\layoutstyle{6x9}

\newcommand{\msun}{\mbox{$M_{\odot}$}}
\newcommand{\be}{\begin{equation}}
\newcommand{\ee}{\end{equation}}
\newcommand{\noi}{\noindent}

\begin{document}

\title{Winds From Massive Stars, And Implications for Supernovae and Gamma-Ray Bursts}


\author{Vikram V. Dwarkadas}{
  address={Astronomy and Astrophysics, Univ of Chicago, 5640 S Ellis Ave, AAC 010c, Chicago, IL 60637},
}



\begin{abstract}
 We review the effects of winds from massive O and B stars on the
 surrounding medium over the various stages of stellar
 evolution. Furthermore we discuss some of the implications for SNe
 and GRB evolution within this wind-blown medium.
\end{abstract}

\maketitle


\section{Introduction}

Massive stars are stars $> 8\msun$ that don't end their lives as white
dwarfs. These stars lose mass during their evolution, with more
massive stars losing a considerable amount of material \cite{WHW:2002}
before they explode as supernovae. This material collects around the
star, forming extended envelopes in some cases, and wind-blown
'bubbles' surrounded by a dense shell in other cases. Herein we review
the development of the circumstellar material around massive stars
during the various phases of evolution. We then explore the impact of
this material on the subsequent evolution of the SN shockwave, and
possibly GRB blast wave, upon the death of the star.

\section{Winds From Massive Stars}

The evolution of massive stars depends on several factors, the main
ones being the initial mass of the star, its rotation velocity, the
presence or absence of a binary companion(s), and the metallicity. A
single, non-rotating massive star at solar metallicity starts its life
as an early type main-sequence O/ B star. These stars lose mass via
fast (few thousand km/s) and dense (10$^{-8} - 10^{-6}~\msun$
yr$^{-1}$) radiatively driven winds \cite{KP:2000}. The interaction of
these winds with the surrounding medium gives rise to ``wind-blown
bubbles'' (WBBs) surrounding the star, bordered by a dense shell. For
a wind with fixed parameters (velocity, mass-loss rate), the density
structure with radius shows 4 different regions (1) a freely expanding
wind with density decreasing as r$^{-2}$, (2) a region of shocked wind
(3) a region of shocked ambient material and (4) the unshocked ambient
medium. Regions (1) and (2) are separated by an inwardly moving (in a
Lagrangian sense) wind-termination shock, (2) and (3) by a contact
discontinuity and (3) and (4) by an outwardly expanding shock
wave. The presence of evolving winds with changing parameters, the
growth of hydrodynamic instabilities and the onset of turbulence will
all cause substantial changes in the structure and morphology of the
bubble.

The radius of the outer shock of the bubble (the outer edge of the
dense shell) is \cite{wmc77}:

\be
\label{eq:rbub}
R_{sh} = 0.76 ~ \left( {L \over \rho} \right)^{1/5} t^{3/5}
\ee

\noi
where $L = 0.5 \dot{M} v_{w}^2 $ is the mechanical luminosity of the
wind, $\dot{M}$ is the wind mass-loss rate and $v_{w}$ is the wind
velocity. In the case of a main-sequence (MS) star with $\dot{M} =
10^{-7} \msun$ yr$^{-1}$, and velocity 2500 km s$^{-1}$, $L$ is about
1.984 $\times$ 10$^{35}$ ergs s$^{-1}$. If this lasts for about 10
million years, the total energy released will be about

\be E_{ms} = 6.25 \times 10^{49}\; {\dot{M}_{-7}}\; v^2_{2500} \; {t_{10}}
~~~ {\rm ergs}
\ee

\noindent
where ${\dot{M}_{-7}}$ is the mass-loss rate in terms of
10$^{-7}\msun$ yr$^{-1}$, ${v_{2500}}$ is the wind velocity in units
of 2500 km s$^{-1}$ and time is in units of 1 million years years
($t_{10}$ = 10 $\times 10^6$ years).

We assume that the main sequence star is formed in a medium with an
average density of about 2.34 $\times$ 10$^{-23}$ g cm$^{-3}$ (a
number density $\sim 10$ particles cm$^{-3}$, appropriate for an ionized
region). From equation \ref{eq:rbub} the radius of the swept-up shell
will be

\be
\label{eq:rbubms}
R_{ms} = 48.8 ~{\dot{M}_{-7}^{1/5}} \; v^{2/5}_{2500} \; {\rho}_{10}^{-1/5} \;
{t_{10}}^{3/5}~~ {\rm pc}
\ee

\noi
In some cases the bubble may reach pressure equilibrium with the
surrounding medium and stall before attaining this radius. Given the
large radius and swept-up mass of the MS shell, the subsequent
evolution is more or less confined to the MS cavity. The swept-up mass
lies in a thin shell surrounding the cavity, which consists of an
inner wind region followed by a region of almost constant density. The
total wind mass ejected over time $t$ is $\dot{M} t$. The mass in the
freely expanding wind is $\dot{M} R_t/v_w$ ($R_t$ = radius of wind
termination shock). Since $v_w ~t >> R_t$ by the end of the MS phase,
a lower limit to the average cavity density is obtained by assuming
that the wind material is uniformly distributed:

\be
\label{eq:bubden}
{\rho}_{bub} = {3 \dot{M} t \over 4 \pi R_{sh}^3} = {3 \over {4 \pi ~ 0.76^3}}
\left( 2 \dot{M}^{2/3} \rho_a \over v_w^2 \right)^{3/5} t^{-4/5}
\ee

\noi
which for the main sequence stage can be written as

\be
\label{eq:bubdennum}
{\rho}_{bubMS} = 1.5 \times 10^{-28}~ {\dot{M}_{-7}^{2/5}}\;
{\rho}_{10}^{3/5} \; v_{3000}^{-6/5}~ {t_{10}}^{-4/5}~~ {\rm g~ cm^{-1}}
\ee

Although this is a lower limit, especially if the bubble stalls early
in the MS phase, it can be seen that the density in the interior of
the bubble is on the order of $10^{-4}-10^{-3} $ particles cm$^{-3}$,
orders of magnitude lower than that of the surrounding medium.

The main-sequence stage is followed by an intermediate stage. For
stars with M $ \leq 50 \msun$ this is generally a red supergiant
(RSG). These stars have large envelopes and slow winds (10-20 km
s$^{-1}$) with a high mass loss rate of 10$^{-5}$ to 10$^{-4}~\msun$
yr$^{-1}$. This creates a high density region around the star,
confined by the pressure of the main-sequence bubble. For a RSG
lifetime of about 200,000 years the total energy input is

\be
E_{RSG} = 8 ~\times 10^{46} ~\dot{M}_{-4} ~v^2_{20} ~ t_{0.2} ~~ {\rm
ergs}
\ee

\noindent
where $\dot{M}_{-4}$ is the mass loss rate in terms of 10$^{-4}
{\msun}$ yr$^{-1}$, v$_{20}$ is the velocity in units of 20 km
s$^{-1}$, and $t_{0.2}$ is the time in units of 200,000 years. The RSG
wind, with its low velocity, should be able to expand a distance
$R_{RSG}$, with wind density $\rho_{RSG}$

\be
R_{RSG} = \kappa ~ 4.2 ~v_{20}~ t_{0.2}~ {\rm pc};~~~~~~~~~\rho_{RSG}
= 2.81 \times 10^{-23}  ~\dot{M}_{-4}~v_{20}^{-1}~ r_{pc}^{-2}
\ee

\noi where we have added a factor $\kappa \ge 1 $ to account for the fact
that neither the transition, nor the change in velocity, from the MS
to the RSG phase is instantaneous. Given the size of the MS bubble, it
is clear that the RSG wind region will be confined to a small fraction
of the main-sequence bubble. The total mass lost during the RSG phase
is

\be
M_{RSG} = 20 ~ \dot{M}_{-4} ~ t_{0.2} ~~ \msun
\ee

Thus although total energy of the outflow is small compared to the MS
stage and the subsequent Wolf-Rayet stage, a large amount of stellar
mass is lost in the RSG stage.

Stars less than about 35$\msun$ will end their lives as red
supergiants and explode to form Type II supernovae (SNe). Stars larger
than this but less than about 50 $\msun$ will lose their outer
envelopes and form Wolf-Rayet (W-R) stars after the RSG
phase. According to some models \cite{LHL:1994} stars $ > 40~\msun$
may become unstable for a short-period of time and pass through a
Luminous Blue Variable (LBV) phase, in which they may lose much larger
amounts of mass in a very short time. The extreme example is $\eta$
Carinae, which is supposed to have lost 2-10 $\msun$ of material in
about 20 years. These may form LBV windblown nebulae within the
main-sequence nebula. Many are known to be bipolar.

Solar metallicity stars above about 35 $\msun$ end their lives as W-R
stars. The mass-loss decreases somewhat to about $10^{-5} {\msun}$
yr$^{-1}$, while the wind velocity increases to 2000 km s$^{-1}$. For
a lifetime of 100,000 years, the total energy input in the W-R phase
is then

\be
E_{WR} = 4 \times 10^{49}~ \dot{M}_{-5}~ v^2_{2000} ~t_{0.1} ~{\rm ergs}
\ee

Although the total mass is less, W-R winds may posses enough momentum
to push out, and possibly break up, any dense shell surrounding the
star from the previous intermediate wind stage, distributing its
contents throughout the surrounding medium. Generally they will have
enough momentum to collide with the MS shell, sending a reflected
shock back. A W-R wind termination shock will be formed where the
thermal pressure of the shocked wind bubble equals the ram pressure of
the freely flowing wind.

The post-MS stages may add considerable mass to the bubble without
increasing the volume much. However equations (\ref{eq:bubden}) and
(\ref{eq:bubdennum}) imply that, even increasing the mass ($\dot{M}
t$) by a factor of 30-40 results in number densities of order 10$^{-3}
- 10^{-2}$ cm$^{-3}$. Therefore the density over the bubble interior
is in general low for W-R bubbles.

Multidimensional calculation \cite{vvd06a, vvd06b, fhy06, glm96}
reveal the presence of hydrodynamic instabilities in many stages. In
one calculation of a 35 $\msun$ star, \cite{vvd06b} found that both
the RSG and subsequent W-R wind (expanding into the RSG wind) were
Rayleigh-Taylor unstable (Fig 1a). These instabilities tend to
break-up the RSG shell and distribute its material over the entire
wind-blown bubble. They may lead to the formation of blobs, clumps and
filaments in the WBB. Eventually, due to density fluctuations in
various stages, the bubble interior becomes quite turbulent by the end
of the simulation.

Other factors will considerably modify this simple picture. Rotation
of the star can lead to enhanced mass-loss \cite{mm00}. Also a star
rotating close to its break-up velocity may emit a wind preferentially
in the polar direction \cite{do02}. Mass-loss rates are also
metallicity dependent, although rotation can lead to high rates even
at low metallicity. And a binary companion can significantly alter the
evolution.

\begin{figure}
\includegraphics[height=.32\textheight]{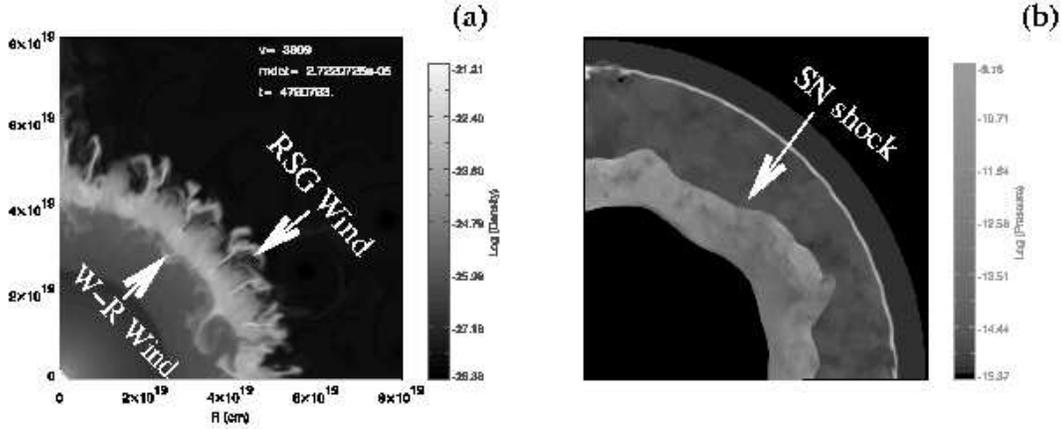} 
\caption{(a)
  The evolution of the W-R wind into the RSG wind region in a 35
  $\msun$ star. Both winds are R-T unstable. (b) The SN shock wave
  expanding within the turbulent interior of a wind-blown bubble can
  becomes wrinkled and loses its spherical shape. For further details
  see \cite{vvd06b}.}
\end{figure}

\section{Implications for Supernovae}

The star will end its life in a SN explosion, giving rise to a shock
wave that expands outwards within this medium. Salient features of the
expansion of the shock wave within the ambient medium have been
revealed by numerical simulations \cite{tbf90, vvd05, vvd06b}:\\
\noindent
$\bullet$ Massive stars will generally be surrounded by low-density bubbles, 
whose size is set by the main sequence stage. If there is a RSG or LBV
phase then the density may be much higher closer in to the
star. However a subsequent W-R phase may destroy this high density
medium and redistribute the material within the bubble.\\
$\bullet$ In either case, the medium near the star is a freely flowing
wind, whose density varies as r$^{-2}$ if the wind parameters are
constant. If the parameters are not constant then the density
variation may be different. In SN 1993J there are several claims
\cite{sn95, mdb01, pam01, bbr02} that the density of the medium
immediately surrounding the star drops less steeply than r$^{-2}$,
with some indications that the density exponent could be as low as
r$^{-1.4}$.\\ $\bullet$ The energy deposited by the winds ($\sim
10^{49} - 10^{50}$ ergs, depending on the initial mass of the star)
into the surrounding medium is a small but non-negligible fraction of
the total SN kinetic energy ($\sim 10^{51}$ ergs).\\ $\bullet$
Supernovae arising from RSG stars will first expand within the
high-density wind before expanding in the low-density cavity. SNe
arising from W-R stars will expand in the freely expanding W-R wind
before encountering the low-density bubble.\\ $\bullet$ Since the
emission from the remnant after the first few months is mainly due to
circumstellar interaction \cite{cf94}, and depends on the density, the
low density implies that the emission will be considerably reduced in
optical, X-rays and radio. Thus SNe in bubbles show much lower
luminosities than their counterparts in the ISM.\\ $\bullet$ Since the
mass is mainly contained in the dense shell bordering the bubble, the
further evolution of the shock wave depends on the dense
shell. Numerical calculation confirm \cite{tbf90, vvd05} that the
expansion depends basically on one parameter $\Lambda =$ mass of
shell/mass of ejecta. If $\Lambda \ge 1$ then the shell begins to have
a significant effect. For low values of $\Lambda$ the shock wave
forgets about the interaction over a period of a few doubling
times. But as $\Lambda$ increases the kinetic energy of the ejecta is
more rapidly converted to thermal energy within the shell, which may
be radiated away. A transmitted shock enters the shell, while a
reflected shock expands back into the ejecta towards the center.  \\
$\bullet$ The interaction of the shock with the high-density shell
leads to a compression of the shell, increasing its density and and
therefore the optical and X-ray emission.\\ $\bullet$ 2D simulations
show that the SN shock expanding in the turbulent interior may not
remain spherically symmetric but acquires a corrugated shape (Fig 1b),
due to its interaction with density fluctuations within the
interior. The impact of the shock with the dense shell occurs in a
piecewise fashion, with different parts of the shock colliding with
the shell at different times. The emission from different regions of
the shell increases in luminosity at different times.  Correspondingly
the reflected shock does not reach the center as one contiguous piece
but as several smaller shocks at different times.\\

\section{Some Thoughts about Gamma-Ray Bursts}

Gamma-Ray Bursts (GRBs) are now widely thought to originate from
Wolf-Rayet stars. The blast wave will then evolve in the wind-blown
bubble surrounding the W-R star, as described above. It should
therefore be possible to explain absorption line spectra seen in GRBs
in the context of this model. The absorption line spectra indicate
multiple absorption systems spanning (roughly) three velocity ranges
(although not all GRBs show all velocity features) - high-velocity
(1000-3000 km/s), intermediate velocity (few hundred km/s), and low
velocity (tens of km/s). The first could arise in the freely expanding
wind, the last slow velocity feature in the slowly-moving dense
shell. The dense shell could also account for the high column density
of neutral hydrogen (up to N$_H > 10^{22}$) seen in some GRBs
\cite{bpc06}. The intermediate features are somewhat more problematic,
because the temperature in the shocked wind is too high for much
absorption to take place. They have been explained by
\cite{vlg05} as due to fragmented W-R and RSG shells. In their analysis
\cite{vlg05} did not consider the various individual ionization
species, the detailed temperature structure, or the presence of an
optical flash. These were taken into account by \cite{lpf06} to
explain the high-velocity absorption line spectra of GRB 021004. Using
a time-dependent photo-ionization code, these authors determined that,
since most of the material within a few parsecs of the star is fully
ionized, the wind-termination shock in GRB021004 must lie at a
distance $> 100$ pc to explain the high-velocity absorption\footnote{A
similarly large distance for the absorption feature in other GRBs was
found by \cite{pcb06}}, assuming it is local to the GRB. This implies
a very large WBB stretching many hundreds of parsecs, which (see
eq.~\ref{eq:rbubms}) therefore must have arisen in a medium with a
very low number density of around 10$^{-3} - 10^{-2}$ cm$^{-3}$.

Several observations of GRB afterglows suggest that the burst is
occurring in a constant density medium \cite{pk02}. In the above
picture, the only relatively constant density medium exists beyond the
wind termination shock. Therefore it implies that the wind-termination
shock is very close-in to the star \cite{clf04}, at a distance of
perhaps 1000's of AU rather than parsec scales. One way to obtain such
a structure is if the density of the ambient medium is extremely high,
in effect compressing the bubble outlined above to a very small
size\footnote{A high pressure without a high density may work, but
requires unrealistically high temperatures}. Our simulations show that
densities $> 10^4$ particles cm$^{-3}$ would be required to achieve
this. Thus some afterglow observations imply very high ambient
densities, whereas absorption line spectra sometimes indicate very low
densities, spanning a total ambient number density range of 7 orders
of magnitude ($10^{-3}-10^4$). If this scenario is correct, then GRB
progenitor stars are born (or at least spend a good fraction of their
lives) in regions of both very high and very low density. It could
indicate that more than one progenitor family is present. If the
density range seems excessively large, then it could suggest that the
absorbing material is not local but in the foreground, and/or our
interpretation of GRB afterglows in a constant density medium is
suspect.


\begin{theacknowledgments}
VVD is supported by award \# AST-0319261 from the National Science
Foundation, and by NASA through grant \# HST-AR-10649 awarded by the
Space Science Telescope Institute. I would like to thank the AAS for
an Intl.~Travel grant that allowed me to attend this meeting, and the
organizers for a splendid conference in a beautiful setting.
\end{theacknowledgments}



\bibliographystyle{aipprocl} 


\end{document}